\documentclass{aastex}
\usepackage{spr-astr-addons}
\usepackage{url}

\RequirePackage{color}

\newcommand{\emaila}{authors@email.com}

\begin{document}

\title{Electron-acoustic solitary structures in two-electron-temperature plasma with
superthermal electrons} \slugcomment{Not to appear in Nonlearned J.,
45.}
%% Running heads
\shorttitle{Electron-acoustic solitary structures}
\shortauthors{Chen and Liu}

\affil{S. Q. Liu (Corresponding author) \\ e-mail:
sqlgroup@ncu.edu.cn}
\author{H. Chen\altaffilmark{}}
\altaffiltext{}{School of Materials Sciences and Engineering,
Nanchang University, Nanchang 330047, P. R. China}
 \and \author{S. Q. Liu\altaffilmark{}}
\altaffiltext{}{School of Sciences, Nanchang University, Nanchang
330047, P. R. China.}

\email{\emaila}

\begin{abstract}
The propagation of nonlinear electron-acoustic waves (EAWs) in an
unmagnetized collisionless plasma system consisting of a cold
electron fluid, superthermal hot electrons and stationary ions is
investigated. A reductive perturbation method is employed to obtain
a modified Korteweg--de Vries (mKdV) equation for the first-order
potential. The small amplitude electron-acoustic solitary wave,
e.g., soliton and double layer (DL) solutions are presented, and the
effects of superthermal electrons on the nature of the solitons are
also discussed. But the results shows that the weak stationary EA
DLs cannot be supported by the present model.
\end{abstract}


%\section*{}
%\label{sec:intro}

\section{Introduction}

The EAWs can either exist in a two temperature (cold and hot)
electron plasma (Watanabe and Taniuti, 1977; Yu and Shukla, 1983) or
in an electron-ion plasma with ions hotter than electrons (Fried and
Gould, 1961). EAWs are typically high-frequency (by comparison with
the ion plasma frequency), dispersive plasma waves where the
relatively cold inertial electrons oscillate against a thermalized
background of inertialess hot electrons which provide the necessary
restoring force. In the long-wavelength approximation, the
dispersion relation of EAWs is given as $\omega = k\left( {{n_{c0} }
\mathord{\left/ {\vphantom {{n_{c0} } {n_{h0} }}} \right.
\kern-\nulldelimiterspace} {n_{h0} }} \right)^{1 \mathord{\left/
{\vphantom {1 2}} \right. \kern-\nulldelimiterspace} 2}\mbox{v}_{th}
$, where $\mbox{v}_{th} = \left( {{\kappa _B T_h } \mathord{\left/
{\vphantom {{\kappa _B T_h } {m_e }}} \right.
\kern-\nulldelimiterspace} {m_e }} \right)^{1 \mathord{\left/
{\vphantom {1 2}} \right. \kern-\nulldelimiterspace} 2}$ is the hot
electron thermal speed, $\kappa _B$ is the Boltzmann constant, and
$n_{c0}(n_{h0}) $ are the cold (hot) equilibrium electrons
densities. The phase speed $c_s = \left( {{n_{c0} } \mathord{\left/
{\vphantom {{n_{c0} } {n_{h0} }}} \right. \kern-\nulldelimiterspace}
{n_{h0} }} \right)^{1 \mathord{\left/ {\vphantom {1 2}} \right.
\kern-\nulldelimiterspace} 2}\mbox{v}_{th} $ of EAWs must be
intermediate between cold and hot electron thermal velocities so
that the Landau damping is avoided. In the Maxwellian plasmas, Gary
and Tokar (1985) performed a parameter survey and found that the hot
electron component constitutes a non-negligible fraction of the
total electron density (more than $\sim $20{\%} ) for the existence
of the EAWs. And for the Lorentzian plasmas, Mace and Hellberg
(1990) showed that the EAWs were usually strongly Landau damped by
the hot electrons, unless ${n_{c0} } \mathord{\left/ {\vphantom
{{n_{c0} } {n_{h0} }}} \right. \kern-\nulldelimiterspace} {n_{h0} }
\ll 1$.

In the nonlinear wave studies, the propagation  of solitary waves is
important as it describes the characteristics of interaction between
waves and plasmas. Solitary waves are localized nonlinear wave
phenomena which arise due to a delicate balance between nonlinearity
and dispersion. Among the best known paradigms used to investigate
small-amplitude nonlinear wave behavior are different versions of
KdV equation (Washimi and Taniuti, 1966), or nonlinear
Schr\"{o}dinger equation (NLSE) (Hasegawa, 1975). Some form of
reductive perturbation technique is used to derive such equations.
The KdV or mKdV describes the evolution of unmodulated wave, while
the NLSE governs the dynamics of a modulated wave packet. In
addition, for the arbitrary amplitude solitary waves, the Sagdeev
pseudopotential method (Sagdeev, 1966) is used too. Electrostatic
solitary structures are often observed in the space and laboratory
plasma environment. EA soliton has been considered as one of the
possible candidates for some of the observed solitary structures.
Recently, the propagation of the linear as well as nonlinear EAWs
has received a great deal of renewed interest not only because the
two electron temperature plasma is very common in laboratory
experiments (Derfler and Simonen, 1969; Henry and Treguier, 1972)
and in space (Dubouloz et al., 1991, 1993; Pottelette et al., 1999;
Berthomier et al., 2000; Singh and Lakhina, 2001), but also because
of the potential importance in interpreting electrostatic component
of the broadband electrostatic noise (BEN) as being solitary EA
structures observed in the cusp of the terrestrial magnetosphere
(Tokar and Gary, 1984; Singh and Lakhina, 2001), in the geomagnetic
tail (Schriver and Ashour-Abdalla, 1989), in auroral region
(Dubouloz et al., 1991, 1993; Pottelette et al., 1999), in the
numerical simulation (Lu, Wang and Dou, 2005; Lu, Wang and Wang,
2005), and in laboratory experiment (Lefebvre, et al., 2011), etc.

On the other side, since Alfv\'{e}n and Carlqvist (1967) had
suggested the current disruption theory for solar flare, the subject
of DL (sometimes also called shock or kinks) has attracted great
attention (Li, 1984; Liu, 2010; reference therein). DLs occur
naturally in a variety of space plasma environments. It turns out
that DLs have the electrostatic potential and other relevant
parameters monotonically changing from one value at one extreme to
another at the other end, hence ``kinks''. This is associated with
adjacent positive and negative charge regions, which give rise to
the name ``double layers". Such DLs are more difficult to generate
and require a fine tuning of the plasma parameters, hence a more
complicated plasma compositions with enough leeway to obey the
necessary constraints (Verheest, 2006; Hellberg, 1992; Moslem,
2007).

The study of nonlinear EAWs has been focused by many authors with
different particle distribution, i.e., Cairns distribution (Pakzad
and Tribeche, 2010), Vortex-like distribution (Mamun and Shukla,
2002), q-nonextensive distribution (Gougam and Tribeche, 2011),
quantum plasmas (Masood and Mushtaq, 2008), et al. And it had been
found that the particles distributions play a crucial role in
characterizing the physics of nonlinear waves.

Recently, the plasma with superthermal particles has gained much
attention. Superthermal electrons are often observed in laboratory,
space, and astrophysical plasma environments, viz., the ionosphere,
auroral zones, mesosphere, lower thermosphere, etc (Pierrard, 2010,
and reference therein). The Kappa functions (Vasyliunas, 1968)
characterized by the spectral index $\kappa $ are found to represent
more suitably the particle's velocity distributions observed in
number of space and astrophysical environment. The Kappa function
may recover the Maxwellian distribution in the limit of $\kappa \to
\infty $ and its mathematical characteristics and physical origin
have recently been addressed by Hau and Fu (2007, the references
within). It is worth noting that some theoretical work focused on
the effects of superthermal particles on different types of linear
and nonlinear collective processes in plasmas. For instance, the
linear properties of plasmas in the presence of a Kappa distribution
with excess superthermal particles have been investigated rather
extensively (Summers and Thorne, 1991; Mace and Hellberg, 1995).

More recently, employing a Sagdeev pseudopotential method, the
nonlinear arbitrary amplitude EAWs were studied, the weak stationary
solitons and DLs were also given by expanding the Sagdeev potential
in small amplitude limit (Sahu, 2010; Younsi and Tribeche, 2010). In
their model, the plasma systems were assumed consisting of cold
fluid electrons, superthermal hot electrons and stationary ions. The
same procedure also was founded for the q-nonextensive distributed
hot electrons plasma system (Gougam and Tribeche, 2011). Baboolal et
al. (1991) have showed that, when ion-acoustic DLs in a plasma with
negative ions are considered, one must be especially careful to
ensure that one's solutions meet the criteria for convergence of the
original expansions. Verheest (1993) has arrived at a similar
conclusion considering DLs in dusty plasmas. Mace and Hellberg
(1993) showed that EA DLs can not be supported in an infinite,
homogeneous, unmagnetized and collisionless plasma system consisting
of cool fluid ions, cold fluid electrons and hot Boltzmann
distributed electrons based on the mKdV model. Some same conclusions
were also can be found (Hellberg, et al., 1992).

Thus, with these ideas in mind, we re-investigate here the small but
finite amplitude solitary structures in such
two-electron-temperature plasma with superthermal electron based on
the mKdV model. One of our objective here is to study the nonlinear
effects of super-thermal distribution of hot electrons on the nature
of the small amplitude solitary waves. Another one is to show
whether the DL solution exist or not. This paper is organized as
follows: in Section 2, the basic set of equations is introduced. In
Section 3, we derive the mKdV equation, and the solutions of both
solitons and DLs are given. Finally, some conclusions and
discussions are given in Section 4.

\section{Basic equation}

We consider a homogeneous system of an unmagnetized collisionless
plasma consisting of a cold electron fluid, and superthermal hot
electrons obeying a Kappa distribution, and ions. It is well known
to us that the linear spectrum of EAWs extends only up to the plasma
frequency of the cold electron population, $\omega _{pc} = \sqrt
{{4\pi n_{c0} e^2} \mathord{\left/ {\vphantom {{4\pi n_{c0} e^2}
{m_e }}} \right. \kern-\nulldelimiterspace} {m_e }} $, so at this
high frequency, the ion population plays the role of a neutralizing
background. For a small---but finite amplitude waves propagate one
dimensionally, such system is governed by the following normalized
equations:
\[\frac{\partial }{\partial t}n_c + \frac{\partial }{\partial x}\left(
{n_c u_c } \right) = 0,\]
\[\frac{\partial u_c }{\partial t} + u_c \frac{\partial u_c }{\partial
x} = \alpha \frac{\partial \varphi }{\partial x},
\]
\begin{equation}
\label{eq1} \frac{\partial ^2\varphi }{\partial x^2} =
\frac{1}{\alpha }n_c + n_h - \left( {1 + \frac{1}{\alpha }} \right),
\end{equation}
\noindent and the super-thermal hot electrons density $n_h $ is
given by (Saini, 2009; Rios, 2010):
\begin{equation}
\label{eq2} n_h = \left( {1 - \frac{\varphi }{\kappa - 3
\mathord{\left/ {\vphantom {3 2}} \right. \kern-\nulldelimiterspace}
2}} \right)^{ - \left( {\kappa - 1 \mathord{\left/ {\vphantom {1 2}}
\right. \kern-\nulldelimiterspace} 2} \right)},
\end{equation}
where $n_{c} (n_{h})$ are cold (hot) electrons normalized densities
to the total unperturbed density $n_0 = n_{c0} + n_{h0} $, $u_c $ is
the velocity of cold electrons normalized to the EA velocity $c_s $,
$\varphi $ is the electrostatic potential to ${\kappa _B T_h }
\mathord{\left/ {\vphantom {{\kappa _B T_h } e}} \right.
\kern-\nulldelimiterspace} e$, respectively. Time and space
variables are normalized, respectively, to the inverse of cold
electron plasma frequency $\omega _{pc} ^{ - 1} $ and the hot
electron Debye length $\lambda _{Dh} = \sqrt {{\kappa _B T_h }
\mathord{\left/ {\vphantom {{\kappa _B T_h } {4\pi n_{h0} e^2}}}
\right. \kern-\nulldelimiterspace} {4\pi n_{h0} e^2}} $. The
parameter $\alpha = {n_{h0} } \mathord{\left/ {\vphantom {{n_{h0} }
{n_{c0} }}} \right. \kern-\nulldelimiterspace} {n_{c0} }$. It should
be noted that the superthermal hot electron component constitutes a
non-negligible fraction of the total electron density, as mentioned
in the introduction.

\section{ The mKdV equation and the solution}
Nonlinear electron-acoustic waves are governed by the full set of
Eqs.(\ref{eq1}) and (\ref{eq2}). To derive the mKdV equation
describing the behavior of the system for longer times and small but
finite amplitude EAWs, we employ the familiar reductive perturbation
technique (Washimi and Taniuti, 1966; Mace and Hellberg, 1993). From
the usual considerations of the small-wavenumber dispersion relation
for electron-acoustic waves, we introduce the slow stretched
coordinates: $\xi = \varepsilon \left( {x - \lambda t} \right),\tau
= \varepsilon ^3t$, where $\varepsilon $ is a small dimensionless
expansion parameter and $\lambda $ is the wave speed normalized by
$c_s $. All physical quantities appearing in Eq.(\ref{eq1}) are
expanded as a power series in $\varepsilon $ about their equilibrium
values as:
\begin{equation}
\label{eq3} \left( {{\begin{array}{*{20}c}
 {n_c } \hfill \\
 {u_c } \hfill \\
 \varphi \hfill \\
\end{array} }} \right) = \left( {{\begin{array}{*{20}c}
 1 \hfill \\
 0 \hfill \\
 0 \hfill \\
\end{array} }} \right) + \varepsilon \left( {{\begin{array}{*{20}c}
 {n_c^{\left( 1 \right)} } \hfill \\
 {u_c^{\left( 1 \right)} } \hfill \\
 {\varphi ^{\left( 1 \right)}} \hfill \\
\end{array} }} \right) + \varepsilon ^2\left( {{\begin{array}{*{20}c}
 {n_c^{\left( 2 \right)} } \hfill \\
 {u_c^{\left( 2 \right)} } \hfill \\
 {\varphi ^{\left( 2 \right)}} \hfill \\
\end{array} }} \right) + \cdots.
\end{equation}

We impose the boundary conditions that as $\left| \xi \right| \to
\infty , n_c = 1, u_c = 0,\varphi = 0$. Substituting Eq.(\ref{eq3})
into the system of Eq.(\ref{eq1}), and equating coefficients of
different powers of $\varepsilon $, from the lowest-order equations
in $\varepsilon $, the following results are obtained:
\begin{equation}
\label{eq4} n_c ^{\left( 1 \right)} = - \frac{\alpha \varphi
^{\left( 1 \right)}}{\lambda ^2},u_c ^{\left( 1 \right)} = -
\frac{\alpha \varphi ^{\left( 1 \right)}}{\lambda }.
\end{equation}
Eqs.(\ref{eq2}) and (\ref{eq4}) as well as Poisson equation give the
linear dispersion relation as
\begin{equation}
\label{eq5} \lambda = \sqrt {\frac{2\kappa - 3}{2\kappa - 1}} .
\end{equation}

From the next order of $\varepsilon $, with the aid of equation
(\ref{eq4}), we obtain the following equations:
\begin{equation}
\label{eq6} n_c ^{\left( 2 \right)} = \frac{3\alpha ^2}{2\lambda
^4}\left( {\varphi ^{\left( 1 \right)}} \right)^2 - \frac{\alpha
}{\lambda ^2}\varphi ^{\left( 2 \right)},
\end{equation}
\begin{equation}
\label{eq7} u_c ^{\left( 2 \right)} = \frac{\alpha ^2}{2\lambda
^3}\left( {\varphi ^{\left( 1 \right)}} \right)^2 - \frac{\alpha
}{\lambda }\varphi ^{\left( 2 \right)},
\end{equation}
Poisson equation with the help of Eq. (\ref{eq2}) at $O\left(
{\varepsilon ^2} \right)$ yields
\begin{equation}
\label{eq8} A\left( {\varphi ^{\left( 1 \right)}} \right)^2 = 0;
\quad A = \frac{3\alpha }{2\lambda ^4} + \frac{\left( {2\kappa - 1}
\right)\left( {2\kappa + 1} \right)}{2\left( {2\kappa - 3}
\right)^2}.
\end{equation}
Since we assume that $\varphi ^{\left( 1 \right)} \ne 0$, it follows
that at least $\left| A \right| \sim O\left( \varepsilon \right)$.
Hence it should be included in the next higher order, i.e., $O\left(
{\varepsilon ^3} \right)$ of Poisson equation. Collecting the third
order of $\varepsilon $, that in $O\left( {\varepsilon ^3} \right)$,
from equations (\ref{eq1}) and (\ref{eq2}) using
(\ref{eq4})-(\ref{eq7}), we obtain
\begin{eqnarray}
\label{eq9} \frac{\partial }{\partial \xi }n_c ^{\left( 3 \right)} =
\frac{\partial }{\partial \tau }\left( { - \frac{2\alpha }{\lambda
^3}\varphi ^{\left( 1 \right)}} \right) +  \nonumber
\\ \frac{\partial }{\partial \xi }\left( { - \frac{5\alpha
^3}{2\lambda ^6}\left( {\varphi ^{\left( 1 \right)}} \right)^3 +
\frac{3\alpha ^2}{\lambda ^4}\varphi ^{\left( 2 \right)}\varphi
^{\left( 1 \right)} - \frac{1}{\lambda ^2}\alpha \varphi ^{\left( 3
\right)}} \right),
\end{eqnarray}

\begin{eqnarray}
\label{eq10} n_h ^{\left( 3 \right)} = \frac{2\kappa - 1}{2\kappa -
3}\varphi ^{\left( 3 \right)} + \frac{\left( {2\kappa - 1}
\right)\left( {2\kappa + 1} \right)}{\left( {2\kappa - 3}
\right)^2}\varphi ^{\left( 2 \right)}\varphi ^{\left( 1 \right)} +
\nonumber \\
\frac{\left( {2\kappa - 1} \right)\left( {2\kappa + 1} \right)\left(
{2\kappa + 3} \right)}{6\left( {2\kappa - 3} \right)^3}\left(
{\varphi ^{\left( 1 \right)}} \right)^3.
\end{eqnarray}

Eliminate the third-order perturbed quantities $n_c ^{\left( 3
\right)}$, $u_c ^{\left( 3 \right)}$, and $\varphi ^{\left( 2
\right)}$, we obtain the following mKdV equation for the first-order
perturbed potential:
\begin{equation}
\label{eq11} \frac{2}{\lambda ^3}\frac{\partial }{\partial \tau
}\varphi + 3A_2 \frac{\partial }{\partial \xi }\varphi ^2 + 4A_3
\frac{\partial }{\partial \xi }\varphi ^3 + \frac{\partial
^3}{\partial \xi ^3}\varphi = 0,
\end{equation}
with the coefficients read as
\begin{equation}
\label{eq12} A_2 = - \left( {\frac{\alpha }{2\lambda ^4} +
\frac{\left( {2\kappa - 1} \right)\left( {2\kappa + 1}
\right)}{6\left( {2\kappa - 3} \right)^2}} \right),
\end{equation}
\begin{equation}
\label{eq13} A_3 = \frac{5\alpha ^2}{8\lambda ^6} - \frac{\left(
{2\kappa - 1} \right)\left( {2\kappa + 1} \right)\left( {2\kappa +
3} \right)}{24\left( {2\kappa - 3} \right)^3}.
\end{equation}
In Eq.(\ref{eq11}), $\varphi $ is used in place of $\varphi ^{\left(
1 \right)}$ for brevity. Let us introduce the variable, $\eta = \xi
- M_0 \tau $, where $\eta $ is the transformed coordinates with
respect to a frame moving with velocity $M_0 $, for the steady-state
solution of the mKdV equation (\ref{eq11}). By using the boundary
conditions $\varphi \to 0$ and ${d\varphi } \mathord{\left/
{\vphantom {{d\varphi } {d\eta }}} \right.
\kern-\nulldelimiterspace} {d\eta } \to 0$ at $\left| \eta \right|
\to \infty $, we obtain
\begin{equation}
\label{eq14} \frac{1}{2}\left( {\frac{d\varphi }{d\eta }} \right)^2
+ V\left( \varphi \right) = 0,
\end{equation}
\noindent where $ V\left( \varphi \right) $ is the Sagdeev
pseudopotential (Sagdeev, 1966), reads as
\begin{equation}
\label{eq15} V\left( \varphi \right) = A_1 \varphi ^2 + A_2 \varphi
^3 + A_3 \varphi ^4,
\end{equation}
\noindent with $A_1 = { - M_0 } \mathord{\left/ {\vphantom {{ - M_0
} {\lambda ^3}}} \right. \kern-\nulldelimiterspace} {\lambda ^3}$.
It should be noted that when $M_0 \ll \lambda $, the result of Sahu
(2010), i.e, in which Eq.(18) is recovered for the small amplitude
solitary waves by Sagdeev pseudopotential approach.
\subsection{Solitons solution}
If one neglects the term corresponding to $\varphi ^4$ in the
Sagdeev pseudopotential (\ref{eq15}), the solution of
Eq.(\ref{eq14}) is
\begin{equation}
\label{eq16} \varphi = \varphi _m \sec h^2\left( {\frac{\eta
}{\Delta }} \right),
\end{equation}
\noindent where the soliton peak amplitude $\varphi _m $ and width
$\Delta $ are given by $\varphi _m = - {A_1 } \mathord{\left/
{\vphantom {{A_1 } {A_2 }}} \right. \kern-\nulldelimiterspace} {A_2
}$, $\Delta = 2 \mathord{\left/ {\vphantom {2 {\sqrt { - 2A_1 } }}}
\right. \kern-\nulldelimiterspace} {\sqrt { - 2A_1 } }$. Solution
(\ref{eq16}) represents a small-amplitude stationary EAWs provided
$A_1 < 0$ or $M_0
> 0$, which means that the solitary waves are supersonic, in agreement with
the large amplitude case. It is clear that the nature of the
solitary waves, i.e., whether the system will support compressive or
rarefactive solitary waves, depends on the sign of $A_2 $. If $A_2 $
is positive (negative) a compressive (rarefactive) solitary wave
exists. In our present case, $A_2 < 0$ since $\kappa > 3
\mathord{\left/ {\vphantom {3 2}} \right. \kern-\nulldelimiterspace}
2$ for a physically realistic thermal speed (Summers and Thorne,
1991; Mace and Hellberg, 1995), so that here we would have a
rarefactive soliton.

\subsection{Double layer solution}
For the DL solution, the Sagdeev potential must satisfy the
conditions (Mace and Hellberg, 1993),
 (a): $V\left( \varphi \right) =
0$ at $\varphi = 0$ and $\varphi = \varphi _m $; (b): ${V}'\left(
\varphi \right) = 0$ at $\varphi = 0$ and $\varphi = \varphi _m $ ;
and (c): ${V}''\left( \varphi \right) < 0$ at $\varphi = 0$ and
$\varphi = \varphi _m $. Applying the boundary conditions (a) and
(b) into Eq.(\ref{eq15}), we obtain
\begin{equation}
\label{eq17} \varphi _m = - \frac{A_2 }{2A_3 }, M_0 = - \frac{\left(
{A_2 } \right)^2}{4A_3 }\lambda ^3,
\end{equation}
then Sagdeev pseudopotential (\ref{eq15}) can be written as
\begin{equation}
\label{eq18} V\left( \varphi \right) = A_3 \varphi ^2\left( {\varphi
_m - \varphi } \right)^2.
\end{equation}
The DL solution is then given by (if it exists)
\begin{equation}
\label{eq19} \varphi = \frac{\varphi _m }{2}\left[ {1 - \tanh \left(
{\frac{2\eta }{\Delta }} \right)} \right],
\end{equation}
with $\Delta = {\sqrt {{ - 8} \mathord{\left/ {\vphantom {{ - 8}
{A_3 }}} \right. \kern-\nulldelimiterspace} {A_3 }} }
\mathord{\left/ {\vphantom {{\sqrt {{ - 8} \mathord{\left/
{\vphantom {{ - 8} {A_3 }}} \right. \kern-\nulldelimiterspace} {A_3
}} } {\left| {\varphi _m } \right|}}} \right.
\kern-\nulldelimiterspace} {\left| {\varphi _m } \right|}$
represents the width of the DL provided $A_3 < 0$, i.e.,
\begin{equation}
\label{eq20} \frac{5\alpha ^2}{8\lambda ^6} - \frac{\left( {2\kappa
- 1} \right)\left( {2\kappa + 1} \right)\left( {2\kappa + 3}
\right)}{24\left( {2\kappa - 3} \right)^3} < 0.
\end{equation}

For a DL exists, the condition (\ref{eq20}) must be fulfilled, then
we have that the hot electron concentration $\alpha $ must satisfy
$\alpha < \sqrt{\frac{(2 \kappa +1)(2 \kappa +3)}{15(2 \kappa -
1)^2}} $. It is indicates that $\alpha < 1$ in the both small and
large value of superthermal index $\kappa $. At such large cold
electron density ratio (${n_{c0} } \mathord{\left/ {\vphantom
{{n_{c0} } {n_{h0} }}} \right. \kern-\nulldelimiterspace} {n_{h0} }
> 1)$, linear Landau damping by the cold electrons would become appreciable
(Gary and Tokar, 1985; Mace and Hellberg, 1990). Furthermore, we
note that for this case the mKdV model leads to a value of $\left|
{\phi _m } \right| \sim 1$ in the small superthermal index case and
$\left| {\phi _m } \right| > 1$ in the large superthermal index
case, which is shown in the Fig.1. It clearly does not satisfy the
expanding requirement $\left| {\phi _m } \right| \sim \varepsilon $.
From the aforementioned considerations, we are let to conclude that
Eq.(\ref{eq19}) is not a valid solution of the mKdV equation for
electron-acoustic waves in the present model. So we can draw the
conclusion that the present model may not sustain a double layer
structure. Such a point has also been studied extensively by Mace
and Hellberg (1993) for the two-temperature electrons Maxwellian
plasmas. Indeed, a systematic investigation should be be undertook
to verify this conclusion, but it beyond the scope of present paper.

\section{ Remarks and Conclusion}
To investigate the existence regions and nature of the EA solitary
structures, we have done numerical calculations for different set of
parameters.

Figs.2 and 3 present the effects of the superthermal index $\kappa $
and hot electron concentration ($\alpha )$ on the amplitude and
width of the solitons in the slightly supersonic point. It can be
seen from Figs.2 and 3 that the decreasing the superthermal index
$\kappa $ and increasing in hot electron concentration ($\alpha )$
will decrease the amplitude of soliton.

In summary, we have studied the nonlinear EAWs in an unmagnetized,
collisionless plasma consisting of cold electrons, superthermal hot
electrons, and stationary ions. A reductive perturbation method has
been used to get the mKdV equation which describes the dynamics of
solitons and DLs. The effects of super-thermal index $\kappa $ and
concentration $\alpha $ of the hot electrons to the cold electrons
on the nature of the solitons are also discussed. It is seen that
the effects of the superthermal hot electrons have a very
significant role on the amplitude and width of the weak amplitude EA
solitons. It is also found that the small amplitude DLs cease to
exist in our present model. Considering the wide relevance of
nonlinear oscillations, we stress that the results of the present
investigation should be useful in understanding the nonlinear
features of localized electrostatic acoustic structures in different
regions of the laboratory experiments and space environments.

\acknowledgments We thank Professor X Q Li for the helpful
suggestions and encouragement. This work is supported by the
National Natural Science Foundation of China under the grant
Nos.10963002 and International S{\&}T Cooperation Program of China
(2009DFA02320) and the National Basic Research Program of China (973
Program)(No. 2010CB635112).

The authors are grateful to the referee for his comments for the
improvement of this paper.

\begin{figure}[h!]
\begin{center}
\includegraphics[height=1.8in,width=2.5in]{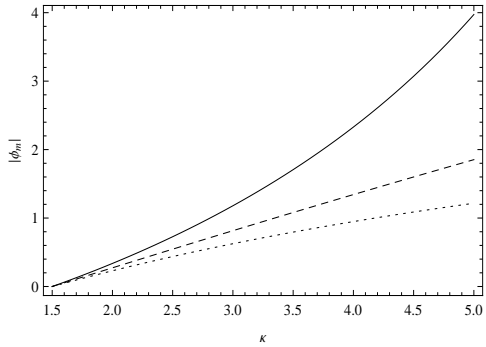}
%% label for entire figure
\caption{Plot of EA DLs amplitude $\left| {\varphi _m } \right|$ vs
$\kappa $, where $\alpha =0.3, 0.25, 0.2$ for solid, dashed, dotted
line, respectively} \label{fig:subfig}
\end{center}
\end{figure}

\begin{figure}[h!]
\begin{center}
\includegraphics[height=1.8in,width=2.5in]{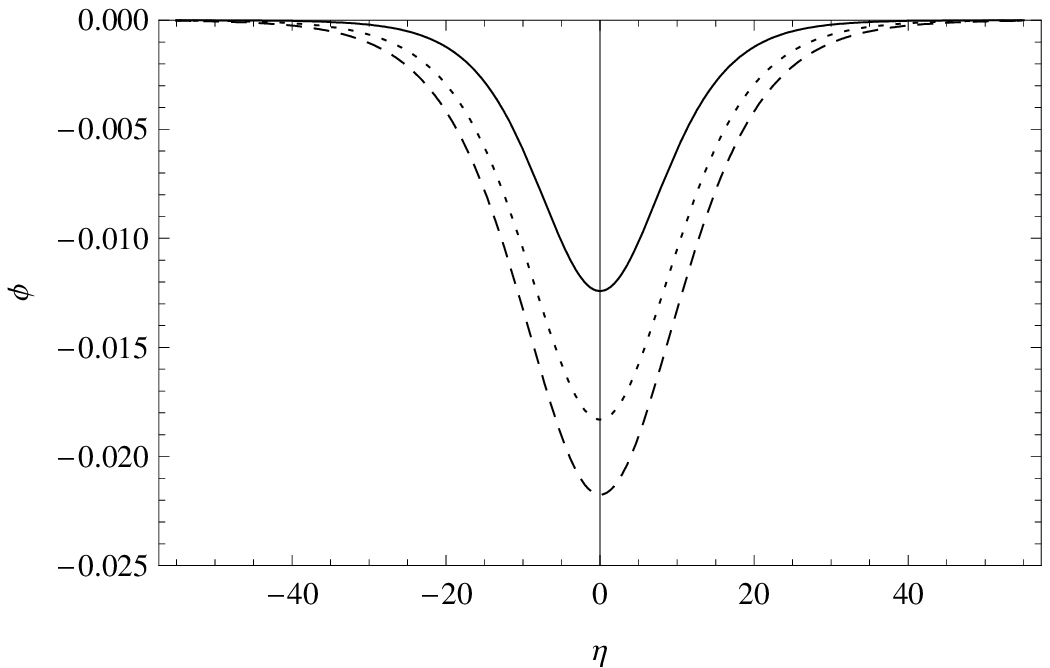}
%% label for entire figure
\caption{Small amplitude soliton for different $\kappa $,\textsf{
}i.e, $\kappa = 3$ for solid line, $\kappa = 6$ for dotted line, and
$\kappa = 15$ for dashed line, where the other parameters are
$\alpha = 0.5$ and $M_0 = 0.1\lambda $. } \label{fig:subfig}
\end{center}
\end{figure}
\begin{figure}[h!]
\begin{center}
\includegraphics[height=1.8in,width=2.5in]{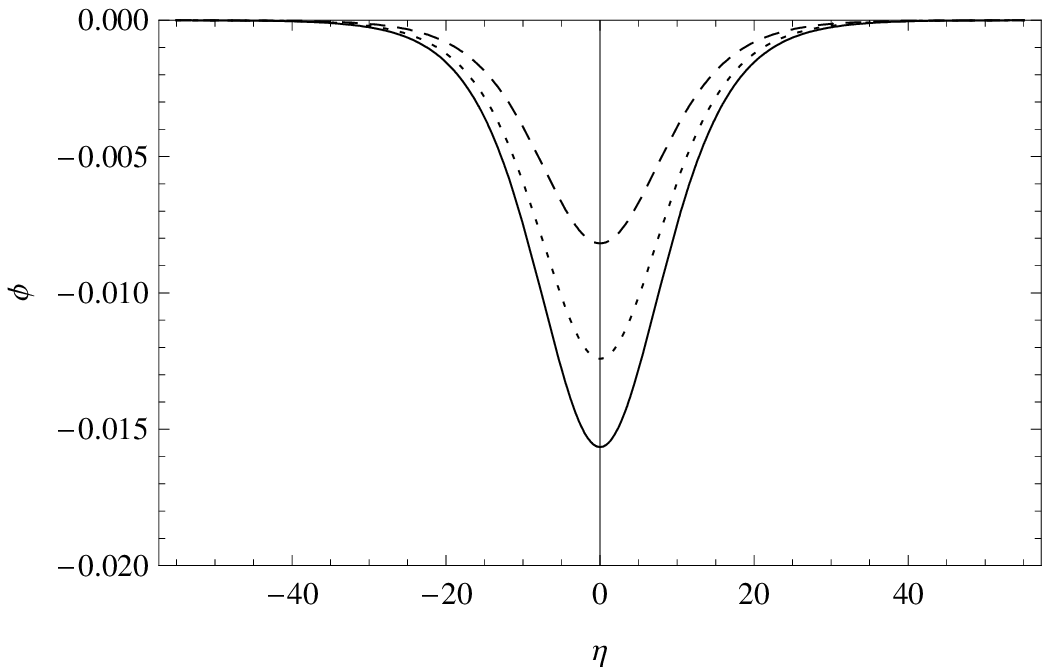}
%% label for entire figure
\caption{Small amplitude soliton for different $\alpha $,\textsf{
}i.e, $\alpha = 0.3$ for solid line, $\alpha = 0.5$ for dotted line,
and $\alpha = 1.0$ for dashed line, where the other parameters are
$\kappa = 3$ and $M_0 = 0.1\lambda $.} \label{fig:subfig}
\end{center}
\end{figure}

%\begin{figure}[h!]
%\begin{center}
%\includegraphics[height=1.8in,width=2.5in]{g3.eps}
%% label for entire figure
%\caption{The allowance hot electron concentration ($\alpha )$ for
%different $\kappa $}\label{fig:subfig}
%\end{center}
%\end{figure}

\end{document}